\documentclass[a4paper,11pt]{iopart}
\pdfoutput=1
\usepackage{iopams}
\usepackage{graphicx}
\usepackage{parskip}

\begin{document}
\title{Finite speed heat transport in a quantum spin chain after quenched local cooling}
\author{Pascal Fries and Haye Hinrichsen}
\address{Universit\"at W\"urzburg, Fakult\"at f\"ur Physik und Astronomie, Am Hubland, \\ 97074 W\"urzburg, Germany}

\ead{pfries@physik.uni-wuerzburg.de, \\ \hspace{13mm}hinrichsen@physik.uni-wuerzburg.de}

\begin{abstract}
We study the dynamics of an initially thermalized spin chain in the quantum XY-model, after sudden coupling to a heat bath of lower temperature at one end of the chain. In the semi-classical limit we see an exponential decay of the system-bath heatflux by exact solution of the reduced dynamics. In the full quantum description however, we numerically find the heatflux to reach intermediate plateaus where it is approximately constant -- a phenomenon that we attribute to the finite speed of heat transport via spin waves.
\end{abstract}

\maketitle

\newcommand{\expe}{\mathrm{e}}
\newcommand{\imi}{\mathrm{i}}
\newcommand{\reals}{\mathbb{R}}
\newcommand{\nats}{\mathbb{N}}
\newcommand{\ints}{\mathbb{Z}}
\newcommand{\Dif}[2][]{{\mathrm{D}^{#1} \mkern-2mu #2 \,}}
\newcommand{\dif}[2][]{{\mathrm{d}^{#1} \mkern-2mu #2 \,}}
\newcommand{\del}[2][]{{\partial^{#1} \mkern-2mu #2 \,}}
\newcommand{\od}[3][]{\frac{\dif[#1]{#2}}{\dif[#1]{#3}}}
\newcommand{\pd}[3][]{\frac{\del[#1]{#2}}{\del[#1]{#3}}}
\newcommand{\md}[3]{\frac{\del[2]{#1}}{\del{#2} \del{#3}}}
\newcommand{\bra}[1]{\left\langle #1 \right|}
\newcommand{\ket}[1]{\left| #1 \right\rangle}
\newcommand{\braket}[2]{\left\langle #1 \middle| #2 \right\rangle}
\newcommand{\ketbra}[2]{\left| #1 \middle\rangle \middle\langle #2 \right|}
\newcommand{\braopket}[3]{\left\langle #1 \middle| #2 \middle| #3 \right\rangle}
\newcommand{\ev}[2]{{\left\langle #1 \right\rangle}_{#2}}
\newcommand{\comm}[2]{\left[ #1, #2 \right]}
\newcommand{\acomm}[2]{\left\lbrace #1, #2 \right\rbrace}
\newcommand{\parens}[1]{{\left( #1 \right)}}
\newcommand{\brackets}[1]{{\left[ #1 \right]}}
\newcommand{\braces}[1]{{\left\lbrace #1 \right\rbrace}}
\newcommand{\lines}[1]{{\left| #1 \right|}}
\newcommand{\eval}[3]{\left.\kern-\nulldelimiterspace#1\vphantom{\big|}\right|_{#2}^{#3}}
\newcommand{\set}[2]{\left\lbrace #1 \middle| #2 \right\rbrace}
\newcommand{\opn}[1]{\mathrm{#1}}
\newcommand{\supop}[1]{\mathcal{#1}}
\newcommand{\order}[1]{{\mathcal{O}\parens{#1}}}
\makeatletter
\newcommand{\const}{\textnormal{const}\@ifnextchar.{}{.}}
\newcommand{\hc}{\textnormal{h.c}\@ifnextchar.{}{.}}
\newcommand{\trace}{\opn{Tr}\@ifnextchar[{\trace@subscr}{}}
    \newcommand{\trace@subscr}[1][]{_{#1} \@ifnextchar[{\trace@arg}{}}
    \newcommand{\trace@arg}[1][]{\brackets{#1}}
\makeatother
\newcommand{\sgn}{\opn{sgn}}

\pagestyle{plain}

\section{Introduction}
If a hot rod is brought in contact with a cold reservoir at one of its ends we expect to see a continuous flow of heat through the boundary until thermal equilibrium is reached. According to our everyday experience, this flux should be proportional to the temperature gradient, implying that it decays exponentially. Not so in the quantum world, where the diffusive character of heat transport, described by Fourier's law, breaks down. On scales up to the free path of phonons, heat transfer is expected to be non-diffusive. Instead, one observes ballistic transport \cite{CasH38}, dominated by a macroscopic drift motion of the phonons, and wave-like \cite{TisL38,LanL41} phenomena, also referred to as ``second sound'', since they show various sound-like features, including a finite propagation velocity, interference phenomena, and reflection at the boundaries of the system \cite{DinR48}. 

In the past, these effects were only found at extremely low temperatures in superfluid Helium \cite{PesV44} and some crystals \cite{AckC68,NarV72}, but recent developements \cite{LeeS15,SieM10} show that nanoscale materials open the way to their observation at high temperatures. Meanwhile, the theory of non-diffusive transport is usually based on consideration of steady state hydrodynamics \cite{CheM63,PhaT13} or constant temperature quantum statistics \cite{BanL13}, leaving unanswered the question to what extent ballistic and wave-like phenomena influence relaxation and thermalization. To study such processes, one has to resort to master equation descriptions, derived in the context of open quantum systems which is, in general, a challenging task. For example, it has been shown \cite{BreH02} that for phonon baths a semi-classical Markovian description of the problem is not sufficient. Instead one has to take the explicit evolution of quantum phases into account and resort to either coarse-graining methods \cite{SchG08} or compensate for a short timespan of non Markovian dynamics \cite{SuaA92,CheY05}. Many questions are still open. Does non-diffusive heat transfer appear in any type of small quantum systems? How does it interact with other features of the system, for example quantum phase transitions? How relevant is the dimension of the system? Is it possible to understand the crossover from the quantum to the macroscopic classical behavior?

In this paper we investigate non-diffusive heat transport in a one-dimensional isotropic quantum XY spin chain. This system was chosen because it is simple, well understood, and can also be solved analytically \cite{LieE61,SonW09}. Moreover, quantum spin chains are known to exhibit spin waves \cite{vKrJ58}, which can be expected to serve as a natural carrier of heat. Although the XY chain does not exhibit a phase transition at finite temperature, its ground state shows a quantum, i.e., zero temperature, phase transition from superfluid to Mott insulating behaviour if the nearest-neighbor interaction is weak enough \cite{SonW09}. In addition, the impact of ballistic transport is expected to be particularly strong in 1D systems \cite{CasH95}. For this reason the chosen model is an excellent candidate for the study of non-conventional heat transport.

We start with the definition of the model and a short summary of known techniques and results in Section~\ref{sec:model}. Our main results are derived and presented in Section~\ref{sec:results}, where we compare different approximation schemes. The paper ends with a discussion in Section~\ref{sec:discussion}. Mathematical details are given in the appendices.

\section{The Model} \label{sec:model}

\subsection{System Hamiltonian}

We consider the isotropic XY-model for a chain of $N$ spin-$\frac 12$ particles in an external field, defined by the Hamiltonian
\begin{eqnarray}
    H &= - \frac j4 \sum_{n=1}^{N-1} \parens{\sigma_n^x \sigma_{n+1}^x + \sigma_n^y \sigma_{n+1}^y} - \frac h2 \sum_{n=1}^N \sigma_n^z \nonumber \\
      &= - \frac j2 \sum_{n \neq N} \parens{\sigma_n^+ \sigma_{n+1}^- + \sigma_n^- \sigma_{n+1}^+} - h \sum_n \sigma_n^+ \sigma_n^- + \const,
\end{eqnarray}
where $\sigma^\pm = \frac 12 \parens{\sigma^x \pm \imi \sigma^y}$ are the spin raising and lowering operators. Following standard techniques \cite{SonW09}, we perform a Jordan-Wigner and Fourier transformation to define the anticommuting (fermionic) fields
\[
    \psi_n = (-\sigma_1^z) \cdots (-\sigma_{n-1}^z) \sigma_n^-, \qquad \tilde \psi_a = \sqrt{\frac 2{N+1}} \sum_{n=1}^N \sin \frac{\pi n a}{N+1} \psi_n,
\]
so that the Hamiltonian attains the diagonal form
\begin{equation}
    H = - \sum_{a=1}^N \omega_a^{} \tilde \psi_a^\dagger \tilde \psi_a^{} + \const. \quad \textnormal{with} \quad \omega_a = h + j \cos \frac{\pi a}{N+1}.
\end{equation}

\subsection{Perturbative master equation}

We now modify the Hamiltonian to couple the first site to an external heat bath via
\[
    H \mapsto H + H_\mathrm{B} + \sigma^x_1 \otimes B,
\]
where $H_\mathrm{B}$ is the bath Hamiltonian and $B$ acts exclusively on the bath. Assuming a weak coupling $\| B \| \ll 1$ as well as a factorizing initial state $\rho_\mathrm{tot}(0) = \rho(0) \otimes \rho_\mathrm{B}$ and some technicalities \cite{SchM07}, we can use second order perturbation theory to obtain the effective master equation
\begin{equation} \label{eq:pert-theory-master-equation-spin-basis}
    \pd{}{t} \rho(t) = - \imi \comm{H}{\rho} - \parens{\int_0^t \dif \tau C(\tau) \comm{\sigma^x_1}{\expe^{-\imi \tau H} \sigma^x_1 \expe^{\imi \tau H} \rho(t)} + \hc},
\end{equation}
where $C(\tau) \equiv \ev{\expe^{\imi \tau H_\mathrm{B}} B \expe^{-\imi \tau H_\mathrm{B}} B}{\rho_\mathrm{B}}$ is the bath auto-correlation function. 

In order to simplify the computation of the interaction term $\int_0^t \dif \tau C(\tau) \expe^{-\imi \tau H} \sigma^x_1 \expe^{\imi \tau H}$, we follow \cite{BreH02,SchG08,SchG11} and switch to the energy eigenbasis
\[
    \ket{k} \equiv \ket{k_1 \ldots k_N} \equiv (\tilde \psi_1^{k_1})^\dagger \cdots (\tilde \psi_N^{k_N})^\dagger \ket{\downarrow \ldots \downarrow}, \qquad E_k = - \sum_a k_a \omega_{k_a}
\]
with $k_a \in \braces{0, 1}$. Introducing the incomplete bath spectral function
\[
    \Gamma_t(\omega) = \int_0^t \dif \tau C(\tau) \expe^{\imi \omega \tau}
\]
and, for the sake of brevity, the notations
\begin{eqnarray*}
    k^{(a)} \equiv (k_1, \ldots, 1 - k_a, \ldots, k_N), \\
    s_{k_a} \equiv 2 k_a - 1, \qquad s^{(a)}_k \equiv (-s_{k_1}) \cdots (-s_{k_{a-1}}), \\
    \textnormal{and} \qquad \hat{\sum_{ab}} \equiv \frac 2{N+1} \sum_{ab} \sin \frac{\pi a}{N+1} \sin \frac{\pi b}{N+1},
\end{eqnarray*}
we can rewrite the above master equation as
\begin{equation} \label{eq:pert-theory-master-equation}
    \pd{}{t} \rho = - \imi  \comm{H + H^-_\mathrm{LS}}{\rho} - \acomm{H^+_\mathrm{LS}}{\rho} + \supop{G} \brackets{\rho},
\end{equation}
with the relaxation generator
\begin{equation} \label{eq:model-generator}
    \supop{G}[\rho] = \hat{\sum_{ab}} \sum_{km} s^{(a)}_k s^{(b)}_{\vphantom{k}m} \parens{\vphantom{\Big()}\Gamma_t (s_{k_a} \omega_a) + \Gamma^*_t (s_{m_b} \omega_b)} \braopket{k^{(a)}}{\rho}{m^{(b)}} \ketbra{k}{m}
\end{equation} 
and the lamb-shift Hamiltonians
\begin{equation} \label{eq:model-lamb-shift}
    H_\mathrm{LS}^\pm = \frac 1{2\sqrt{\pm 1}} \hat{\sum_{ab}} \sum_k s^{(a)}_k s^{(b)}_k \parens{\vphantom{\Big()}\Gamma_t (s_{k_a} \omega_a) \pm \Gamma^*_t (s_{k_b} \omega_b)} \ketbra{k^{(b)}}{k^{(a)}}.
\end{equation}
A detailed derivation of eqns.~(\ref{eq:pert-theory-master-equation})~--~(\ref{eq:model-lamb-shift}) can be found in \ref{sec:generator-form-derivation}.

\subsection{Approximating the incomplete bath spectral function} 

To compute the reduced system dynamics, described by eq.~(\ref{eq:pert-theory-master-equation}), one needs information about the heat bath, encoded in $C(t)$. As was recently shown in \cite{FiaO15}, a bath exhibiting quantum chaos can be effectively described by the spectrum 
\[
    \gamma(\omega) \equiv \Gamma_\infty(\omega) + \Gamma^*_\infty(\omega) = \int_\reals \dif \tau \expe^{\imi \omega \tau} C(\tau) = \lambda^2 \exp \parens{-\frac 12 \parens{\frac\omega\sigma - \frac{\beta \sigma}2}^{\!\!2}},
\]
where $\lambda \propto \|B\|$ is the coupling strength, $\beta \equiv \beta_{\mathrm{bath}}$ is the inverse bath temperature, and $\sigma$ is the inverse decay-timescale of self correlations. To calculate
\[
    \Gamma^{}_t(\omega) = \frac 1{2\pi}  \int_\reals \dif \Omega \gamma(\Omega) \int_0^t \dif \tau \expe^{\imi(\omega - \Omega)\tau},
\]
we use
\[
    \gamma(\omega) = \lim_{n \to \infty} \lambda^2 \parens{1 + \frac 1{2n} \parens{\frac\omega\sigma - \frac{\beta \sigma}2}^{\!\!2}}^{\!\!\!-n}
\]
and the residue theorem around the pole $\frac{\omega_0}\sigma = \frac{\beta\sigma}2 - \sqrt{2 n} \imi$, obtaining
\begin{equation} \label{eq:incomplete-bath-sectrum}
    \Gamma^{}_t(\omega) = \lim_{n \to \infty} \frac {\lambda^2}{4^n} \sum_{k=0}^{n-1} \parens{{2n-2-k}\atop{n-1}} \parens{\frac{\sqrt{8n} \imi \sigma}{\omega-\omega_0}}^{\!\!k+1} \!\!\!\!\! P(k+1, -\imi t (\omega-\omega_0)),
\end{equation}
where
\[
    P(k+1, z) \equiv \frac 1{k!} \int_0^z \dif \tau \tau^k \expe^{-\tau} = 1 - \expe^{-z} \sum_{m=0}^k \frac{z^m}{m!}
\]
is the (lower) regularized gamma function. Provided that $|2\omega - \beta \sigma^2|/\sigma$ is not too large, we can truncate eq.~(\ref{eq:incomplete-bath-sectrum}) at small $n$ and still obtain reasonable results. For the rest of this paper, we choose $n=2$, as higher orders do not seem to cause notable differences.

\subsection{Concatenation scheme and secular approximation}

Because of the explicit time dependence of $\Gamma_t$, the master equation~(\ref{eq:pert-theory-master-equation}) is not local in time. If however the correlation function $C(t)$ decays reasonably fast, we can replace $\Gamma_t \mapsto \Gamma_\infty$ in the long time limit, yielding the \emph{Redfield equation} \cite{SchM07}. This procedure, called Markov approximation, is known to be invalid for short times, since, in general, it violates complete positivity \cite{SuaA92}. To circumvent this problem, we will therefore use the \emph{concatenation scheme} \cite{CheY05} of switching from $\Gamma_t$ to $\Gamma_\infty$ at the transition from non-Markovian to Markovian dynamics, i.e., when $C(t)$ has decayed to an irrelevant value.

In order to see which features of the system are genuinely quantum, we will compare the concatenation scheme with the so called \emph{secular approximation} \cite{BreH02}, a semi-classical approximation with Markovian dynamics on all timescales which is known to guarantee complete positivity \cite{dVeI16}. To this end, note that we can transform eq.~(\ref{eq:pert-theory-master-equation}) into the interaction picture, just by dropping the system Hamiltonian and introducing a phase factor of $\expe^{-\imi (s_{k_a} \omega_a - s_{m_b} \omega_b) t}$ in each summand of eq.\,(\ref{eq:model-generator}) and similarly in eq.~(\ref{eq:model-lamb-shift}). Now, since all terms that oscillate at the rate $\omega_a \pm \omega_b$ of the system's transitions should vanish in the limit $\hbar \to 0$, we use a coarse graining approach and replace
\[
    \expe^{-\imi (s_{k_a} \omega_a - s_{m_b} \omega_b) t} \mapsto \int_\reals \dif \tau \expe^{-\imi (s_{k_a} \omega_a - s_{m_b} \omega_b) (t-\tau)} w(\tau),
\]
where $w$ is a normalized real valued window function with Fourier transform $W(\omega) \equiv \int_\reals \dif \tau \expe^{\imi \omega \tau} w(\tau)$. Back to the Schr\"odinger picture, this results in factors $W(s_{k_a} \omega_a - s_{m_b} \omega_b)$ and $W(s_{k_a} \omega_a - s_{k_b} \omega_b)$ in each summand of eqns.~(\ref{eq:model-generator}) and~(\ref{eq:model-lamb-shift}), respectively. As we will see in the next section, the secular approximation yields results that differ strongly from those of the concatenation scheme -- a phenomenon that is well-known for phonon baths \cite{BreH02} and generally expected in systems where internal and relaxation timescales are comparable.

\section{Results} \label{sec:results}

\subsection{Exponential relaxation in the secular approximation} 

To compute the dynamics of the system in the secular approximation, let us consider the limit of strong coarse graining $W(\omega-\omega') \to \delta_{\omega, \omega'}$ in the Markovian ($\Gamma_t \mapsto \Gamma_\infty$) version of eq.~(\ref{eq:pert-theory-master-equation}): Since the dispersion relation is non-degenerate, we find that the diagonal elements $\rho_{kk} \equiv \braopket{k}{\rho}{k}$ decouple from the other ones, yielding the \emph{rate equations}
\begin{equation}
    \pd{}{t} \rho_{kk} = \frac 2{N+1} \sum_a \sin^2 \frac{\pi a}{N+1} \parens{\vphantom{\Big()}\gamma(s_{k_a} \omega_a) \rho_{k^{(a)} k^{(a)}} - \gamma (-s_{k_a} \omega_a) \rho_{kk}}.
\end{equation}
We can solve them analytically by means of a normalized product ansatz
\[
    \rho_{k k} = \rho^1_{k_1 k_1} \cdots \rho^N_{k_N k_N}, \qquad \rho^a_{11} + \rho^a_{00} = 1,
\]
yielding
\begin{eqnarray}
    \rho^a_{k_a k_a} (t) =  \rho^a_{k_a k_a}(\infty) + \parens{\rho^a_{k_a k_a}(0) -  \rho^a_{k_a k_a}(\infty)} \expe^{-\frac t{\tau_{a}}}, \\
    \rho^a_{k_a k_a}(\infty) \equiv \frac {\gamma(s_{k_a} \omega_a)}{\gamma(\omega_a) + \gamma(-\omega_a)}, \nonumber \\
    \tau_a \equiv \frac{N+1}{2 (\gamma(\omega_a) + \gamma(-\omega_a)) \sin^2 \frac{\pi a}{N+1}} \nonumber,
\end{eqnarray}
i.e., exponential relaxation to a steady state which is thermal iff the \emph{Kubo-Martin-Schwinger condition} $\gamma(\omega) = \expe^{\beta \omega} \gamma(-\omega)$ holds \cite{KosA77,SchG08}.

\begin{figure}[ht]
\begin{center}
    \includegraphics[width=12cm,height=6cm]{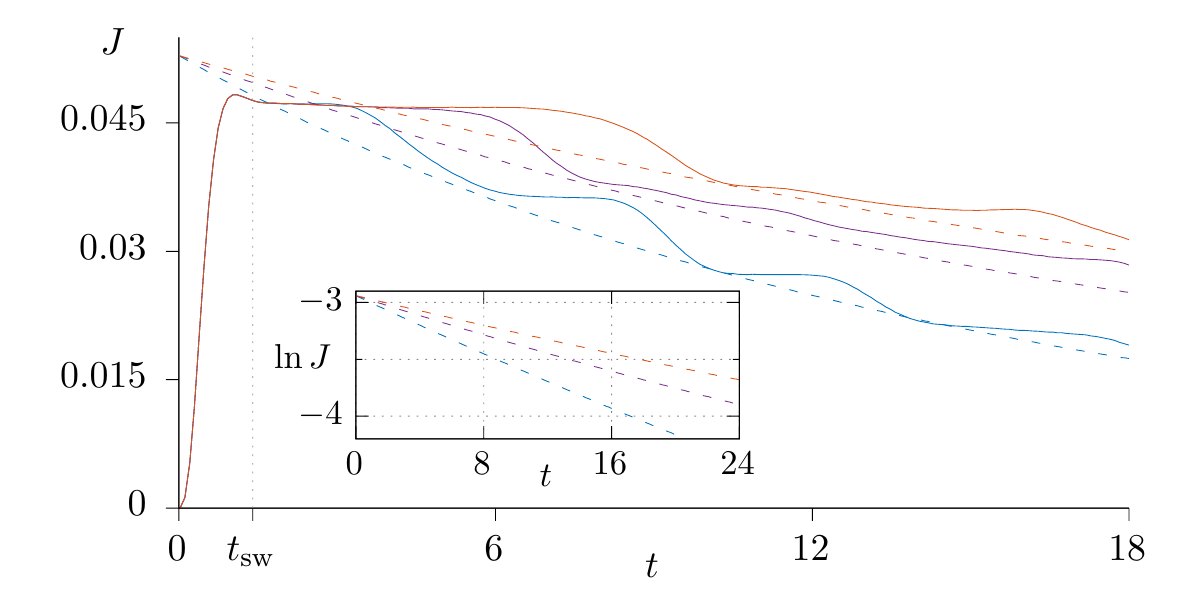}
    \caption{System to bath heatflux $J(t) = -\od{}{t} \ev{H}{\rho(t)}$, according to the concatenation scheme (solid) and the secular approximation (dashed). In the former we switch to the Markov approximation at $t=t_\mathrm{sw}$. The initial state is maximally mixed ($\beta_\mathrm{sys,0} = 0$), the chain length is $N=3,5,7$ (from bottom to top), and other parameters are $h=1$, $j=2$, $\sigma=2.5$, $\beta_\mathrm{bath}=0.8$, $\lambda=0.4$. The leading exponential decay in the secular approximation is shown in the inset.}
    \label{fig:heatflux-comparison}
\end{center}
\end{figure}

We proceed to calculate the system-bath heatflux 
\begin{equation}
    J(t) = -\od{}{t} \ev{H}{\rho(t)} = -\sum_a \frac{\omega_a}{\tau_a} \parens{\rho^a_{11}(0) -  \rho^a_{11}(\infty)} \expe^{-\frac t{\tau_{a}}},
\end{equation}
shown as dashed lines in fig.~\ref{fig:heatflux-comparison}. We see that the heatflux decays exponentially, the timescale being bounded by $\tau_{(N+1)/2}$ and $\tau_1$, i.e.,
\[
\hspace{-1em} \exp\parens{-2t\frac{\gamma(h)+\gamma(-h)}{N+1}} \, \lesssim \, \lines{\frac{J(t)}{J(0)}} \, \lesssim \, \exp\parens{-2t \pi^2\frac{\gamma(h+j)+\gamma(-h-j)}{(N+1)^3}}.
\]

\subsection{Concatenation scheme numerics and heat transport by spin waves}

In order to find out how quantum effects cause deviations from the results of the last subsection, we will now solve the concatenation scheme master equation~(\ref{eq:pert-theory-master-equation}) numerically. 

To this end, note that $\Gamma_t$, as given by eq.~(\ref{eq:incomplete-bath-sectrum}), attains an approximately constant value for $t \sqrt{2n} \sigma \gg 1$, with $n = 2$ being the order of approximation. Hence, we can safely switch from the non-Markovian generator and lamb shift Hamiltonians, as described by eqns.~(\ref{eq:model-generator}) and~(\ref{eq:model-lamb-shift}), to their Markovian form with $\Gamma_t \mapsto \Gamma_\infty$ at $t_\mathrm{sw} = \frac {3.5}\sigma$. In both cases, we use an ordinary Runge-Kutta iteration of order $4$ with adaptive step size. The initial state is taken to be maximally mixed, i.e., thermal with $\beta_\mathrm{sys,0} = 0$.

As can be seen in fig.~\ref{fig:heatflux-comparison} as solid lines, the resulting heatflux $J$ increases rapidly due to buildup of correlations with the bath until it saturates (similar results were found in \cite{SuaA92} for the population in a two level system). Once saturated, $J$ remains approximately constant for a timespan roughly proportional to the chain length $N$, before it decreases to another saturated level. This process repeats itself, getting smeared out at late times in sufficiently long chains. These observations can be made in a wide range of parameters, regardless of the ground state structure given by the specific field strength $|h/j| \lesseqgtr 1$. We argue that this behaviour is caused by the finite speed of heat transport due to spin waves. 

\begin{figure}[ht]
\begin{center}
    \includegraphics[width=14cm,height=6cm]{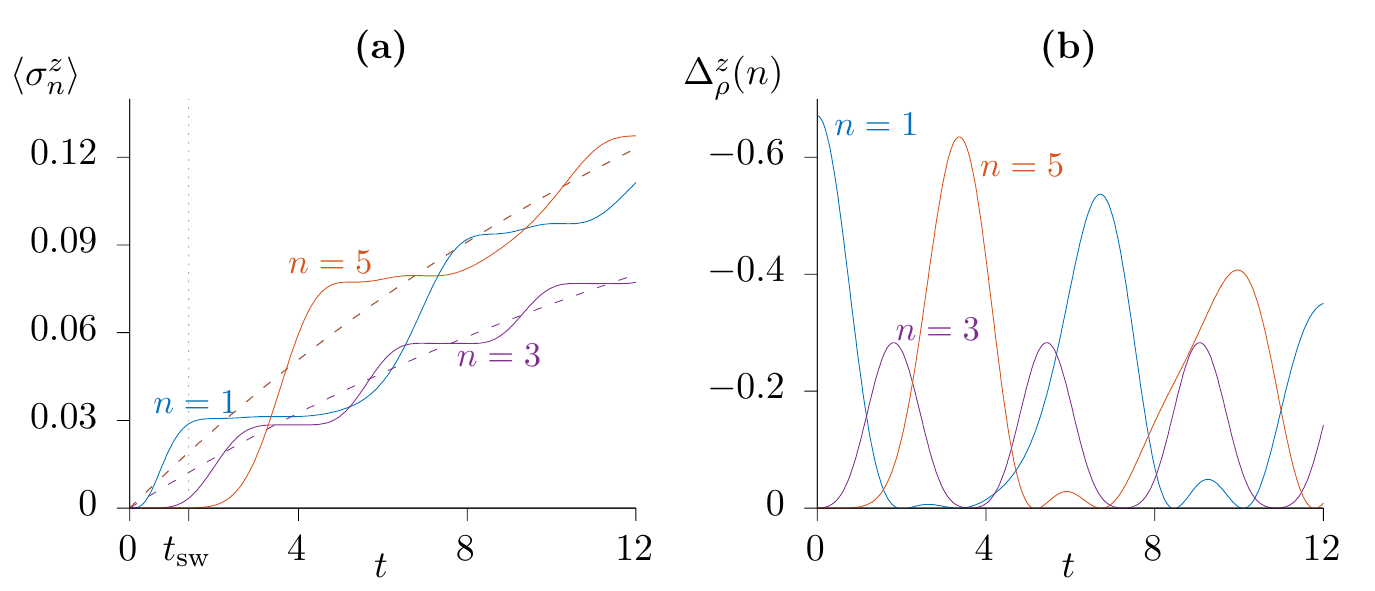}
    \caption{\textbf{(a)} Local magnetization $\ev{\sigma^z_n}{\rho(t)}$, according to the concatenation scheme (solid) and secular approximation (dashed) at sites $n=1,3,5$. The chain length is $N=5$, the initial state and other parameters are the same as in fig.~\ref{fig:heatflux-comparison}. \textbf{(b)} Time dependent response $\Delta^z_\rho(n,t)$ of the local magnetization to a spin flip at site $1$, given by eq.~(\ref{eq:finite-size-spin-flip}) with $N=5$, initial state being thermal with $\beta_\mathrm{sys,0} = 0.8$.}
    \label{fig:spin-waves}
\end{center}
\end{figure}

To see this, consider the local magnetization $\ev{\sigma^z_n}{\rho(t)}$, depicted as solid lines in fig.~\ref{fig:spin-waves}(a). Shortly after the system is brought in contact with the bath, the first site cools down, transferring heat into the bath and aligning itself with the external field. After it reaches a certain threshold, it acts purely as a coupling between the bath and the rest of the chain and the same process is repeated between the first and the second site -- the result being a superposition of spin waves travelling through the chain, which are reflected at the end. As soon as the wave-packet returns to the bath however, the magnetization of the first site will surpass its previous value, causing the heatflux to drop, and reflect again. Again, this phenomenon gets smeared out by dispersion at late times and does not exist in the secular approximation (dashed lines in fig.~\ref{fig:spin-waves}(a)).

\subsection{Unitary dynamics of a single spin flip}

To better understand the propagation of the aforementioned spin waves, let us replace the effect of the heat bath by a single spin flip at site $1$ and study the resulting behaviour. This allows us to derive an analytic expression for the dynamical expectation values
\[
    \Delta^z_k(n,t) \equiv \braopket{k}{\sigma^x_1 \expe^{\imi t H} \sigma^z_n \expe^{-\imi t H} \sigma^x_1}{k} - \braopket{k}{\sigma^z_n}{k},
\]
which describe the time-dependent response of the local magnetization at site $n$ in eigenstate $\ket{k}$. A straightforward but lengthy calculation, which can be found in \ref{sec:spin-flip-derivation}, shows that
\begin{equation} \label{eq:eigenstate-spin-flip}
    \frac{\Delta^z_k(n,t)}2 = \frac 2{N+1} \hat{\sum_{ab}} \expe^{- \imi t (\omega_a - \omega_b)} \sin \frac{\pi n a}{N+1} \sin \frac{\pi n b}{N+1} (1 - k_a - k_b).
\end{equation}

\begin{figure}[t]
\begin{center}
    \includegraphics[width=15cm,height=6cm]{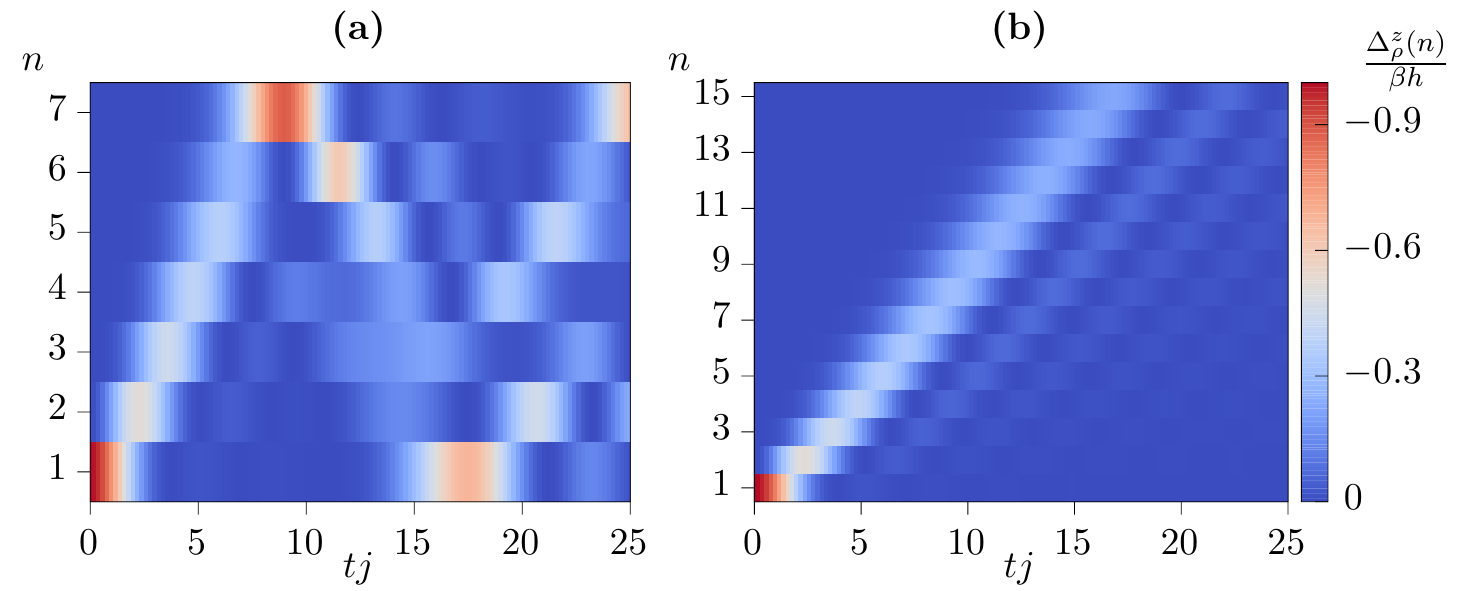}
    \caption{Normalized response of the local magnetization to a spin flip at site $1$ for a thermal initial state of high temperature $\beta_\mathrm{sys,0} \to 0$. \textbf{(a)} Chain of finite length $N = 7$, given by the limit $\beta \to 0$ in eq.~(\ref{eq:finite-size-spin-flip}). \textbf{(b)} Thermodynamic limit $N \to \infty$, described by eq.~(\ref{eq:high-temperature-spin-flip}).}
    \label{fig:spin-flip-image}
\end{center}
\end{figure}

In contrast to the preceeding subsections, we will now assume that the system's state initially describes a canonical ensemble
\[
    \rho = \sum_{k_1} \frac{\expe^{\beta k_1 \omega_1}}{1 + \expe^{\beta \omega_1}} \cdots \sum_{k_N} \frac{\expe^{\beta k_N \omega_N}}{1 + \expe^{\beta \omega_N}} \ketbra{k}{k}
\]
at a finite inverse temperature $\beta \equiv \beta_\mathrm{sys,0} > 0$, since a single spin flip does not alter a fully mixed ensemble at infinite temperature. This amounts to the replacement $(1 - k_a - k_b) \mapsto \frac{1 - \expe^{\beta (\omega_a + \omega_b)}}{(1 + \expe^{\beta \omega_a})(1 + \expe^{\beta \omega_b})}$ in the above expression, causing the sum to factor into
\begin{equation} \label{eq:finite-size-spin-flip}
    \frac{\Delta^z_\rho(n,t)}2 = \lines{\frac 2{N+1} \sum_a \frac {\expe^{\imi t \omega_a}}{1 + \expe^{\beta \omega_a}} \sin \frac{\pi n a}{N+1} \sin \frac{\pi a}{N+1}}^2 - (\beta \mapsto -\beta).
\end{equation}
Plots of $\Delta^z_\rho(n,t)$ and the normalized high temperature limit $\lim_{\beta \to 0} \frac{\Delta^z_\rho(n,t)}{\beta h}$ can be found in figs.~\ref{fig:spin-waves}(b) and~\ref{fig:spin-flip-image}(a), respectively. We clearly see that the spin waves, excited by a single flip at site $1$, travel at the same speed as the change in local magnetization, excited by quenched cooling. Note also that, due to dispersion, the behaviour gets increasingly chaotic at late times. This confirms our expectation that heat transport is related to spin wave propagation.

In the thermodynamic limit $N \to \infty$, we substitute $\omega_a \equiv h + j x$ and find
\[
    \frac{\Delta^z_\rho(n,t)}2  = \lines{\frac 2\pi \int_{-1}^1 \dif x \frac {\expe^{\imi t j x} \sin(n \arccos x)}{1 + \expe^{\beta (h+jx)}}}^2  - (\beta \mapsto -\beta).
\]
Now, for high temperatures $\beta \to 0$, we approximate
\[
    \frac 2{1 + \expe^{\pm \beta (h + jx)}} \sim 1 \mp \beta \frac{h+jx}2
\]
to obtain
\begin{equation} \label{eq:high-temperature-spin-flip}
    \Delta^z_\rho(n,t) \sim - 2 \beta h \parens{\frac{n b_n}{tj}}^{\!\!2}
\end{equation}
with $b_n$ being the Fourier coefficients of the even function
\[
    \sin \parens{tj\cos a - \frac{\pi}4} = \sum_{n=1}^\infty b_n \cos nx + \const.
\]
Note that, in this limit, the nearest neighbour coupling $j$ and the external field $h$ only amount to a rescaling of time and temperature, respectively. A plot of eq.~(\ref{eq:high-temperature-spin-flip}) is shown in fig.~\ref{fig:spin-flip-image}(b). As expected, the response is described by a dispersion wave crest travelling along the chain without reflection.  

On the other hand, in the low temperature regime $\beta \to \infty$, we have
\[
    \frac 2{1 + \expe^{\pm \beta (h + jx)}} \sim 1 \mp \sgn(h+jx),
\]
hence
\begin{equation} 
    \Delta^z_\rho(n,t) \sim -\frac{8\, \sgn(h) n b_n}{\pi tj} \int_{\frac{\pi}2 - \zeta}^{\frac{\pi}2 + \zeta} \dif a \cos \parens{tj\cos a - \frac{\pi}4}  \sin(a) \sin(na)
\end{equation}
with $\zeta \equiv \arcsin \min \parens{1, \lines{\frac hj}}$. Interestingly, in the Mott insulator phase $\lines{\frac hj} \geq 1$, this reduces to eq.~(\ref{eq:high-temperature-spin-flip}) with $\beta \mapsto \frac 2{|h|}$. We therefore expect the dynamics in this regime to be largely temperature independent.

\section{Discussion} \label{sec:discussion}

In this paper we have studied the response of a one-dimensional isotropic quantum XY chain at finite and infinite temperature to a sudden quenched cooling at one of its boundaries. As expected, it turned out that heat is transported by spin waves at a finite velocity. These spin waves exhibit sound-like features, such as reflection and interference. However, spin waves travelling over longer distances loose their integrity, leading eventually to a chaotic behavior which qualitatively explains the crossover to a diffusive type of heat transport.

We find non-diffusive transport to occur at a wide range of interaction strengths, regardless of the ground state being superfluid or Mott-insulating. This is plausible since spin waves occur in both phases.

As a main result, the present work confirms that an accurate master equation description of non-diffusive heat transfer requires a non-Markovian description at short times. Although the dynamics is Markovian at later times, one still has to keep track of the quantum-mechanical phases. To see this, we have also studied the Markovian semi-classical limit, given by the secular approximation, where non-trivial features are lost.

Finally, our study demonstrates that quantum spin chains are suitable candidates for a theoretical study of quantum heat transport. It would be interesting to investigate other systems and situations, in order to access the robustness of the observed phenomena -- especially in higher dimensions, where ballistic and wave-like transport are only observed under special conditions.

\appendix

\section{Generator form of the master equation} \label{sec:generator-form-derivation}

In the energy eigenbasis, eq.~(\ref{eq:pert-theory-master-equation-spin-basis}) has the form
\[
\fl \pd{}{t}\rho = -\imi \comm{H}{\rho} - \parens{\sum_{klmn} \braopket{k}{\sigma^x_1}{l} \braopket{n}{\sigma^x_1}{m} \Gamma_t(E_m - E_n) \comm{\vphantom{\Big()}\ketbra{k}{l}}{\ketbra{n}{m} \rho(t)} + \hc}.
\]
Renaming $k \leftrightarrow m$ and $l \leftrightarrow n$ in the $\hc$-terms, this turns into
\begin{eqnarray*}
\fl \pd{}{t} \rho = -\imi \comm{H}{\rho} - \sum_{klmn} \braopket{k}{\sigma^x_1}{l} \braopket{n}{\sigma^x_1}{m} \\
    \times \Bigg( \Gamma_t(E_m - E_n) \ket{k}\braket{l}{n} \bra{m} \rho + \Gamma^*_t(E_k - E_l) \rho \ket{k}\braket{l}{n} \bra{m} \\
    \hspace{10em} - \parens{\vphantom{\Big()}\Gamma_t(E_m - E_n) + \Gamma^*_t(E_k - E_l)} \ketbra{n}{m} \rho \ketbra{k}{l} \Bigg)
\end{eqnarray*}
and we can expand the sum of products in the second line into commutator and anticommutator, yielding
\begin{eqnarray*}
\fl \pd{}{t} \rho = -\imi \comm{H}{\rho} + \sum_{klmn} \braopket{k}{\sigma^x_1}{l} \braopket{n}{\sigma^x_1}{m} \\
    \times \Bigg( - \frac{1}{2} \parens{\vphantom{\Big()}\Gamma_t(E_m - E_n) - \Gamma^*_t(E_k - E_l)} \comm{\vphantom{\Big()}\ket{k}\braket{l}{n} \bra{m}}{\rho} \\ 
    \hspace{5em} - \frac{1}{2} \parens{\vphantom{\Big()}\Gamma_t(E_m - E_n) + \Gamma^*_t(E_k - E_l)} \acomm{\vphantom{\Big()}\ket{k}\braket{l}{n} \bra{m}}{\rho} \\
    \hspace{10em} + \parens{\vphantom{\Big()}\Gamma_t(E_m - E_n) + \Gamma^*_t(E_k - E_l)} \ketbra{n}{m} \rho \ketbra{k}{l} \Bigg).
\end{eqnarray*}
We thus arrive at eq.~(\ref{eq:pert-theory-master-equation}) with
\[
    \supop{G}[\rho] = \sum_{klmn} \braopket{k}{\sigma^x_1}{l} \braopket{n}{\sigma^x_1}{m} \parens{\vphantom{\Big()}\Gamma_t(E_m - E_n) + \Gamma^*_t(E_k - E_l)} \ketbra{n}{m} \rho \ketbra{k}{l}
\]
and
\[
    H_\mathrm{LS}^\pm = \sum_{klmn} \frac{\braopket{k}{\sigma^x_1}{l} \braopket{n}{\sigma^x_1}{m}}{2\sqrt{\pm 1}} \parens{\vphantom{\Big()}\Gamma_t(E_m - E_n) \pm \Gamma^*_t(E_k - E_l)} \ket{k} \braket{l}{n} \bra{m}.
\]
We can now use the anticommutation relations to calculate
\begin{eqnarray*}
    \braopket{k}{\sigma^x_1}{l} \braopket{n}{\sigma^x_1}{m} 
        &= \braopket{k}{\parens{\psi^{\dagger}_1 + \psi^{}_1}}{l} \braopket{n}{\parens{\psi^{\dagger}_1 + \psi^{}_1}}{m} \\
        &= \hat{\sum_{ab}} \braopket{k}{\parens{\tilde \psi^{\dagger}_a + \tilde \psi^{}_a}}{l} \braopket{n}{\parens{\tilde \psi^{\dagger}_b + \tilde \psi^{}_b}}{m} \\
        &= \hat{\sum_{ab}} s^{(a)}_k s^{(b)}_m \delta_{k^{(a)} l} \delta_{n m^{(b)}},
\end{eqnarray*}
which directly leads to eqns.~(\ref{eq:model-generator}) and~(\ref{eq:model-lamb-shift}).

\section{Response to a single spin flip} \label{sec:spin-flip-derivation}

Let us first consider the time evolution of the action of the spin flip operator
\[
    \sigma^x_1 = \sqrt{\frac 2{N+1}} \sum_{r=1}^N \sin \frac{\pi r}{N+1} \parens{\tilde \psi^\dagger_r + \tilde \psi^{}_r},
\]
on an eigenstate $\ket{k}$, given by
\[
    \expe^{-\imi t H} \sigma^x_1 \ket{k} = \sqrt{\frac 2{N+1}} \sum_r \expe^{\imi t \parens{(1 - k_r)\omega_r + \sum_{j \neq r} k_j \omega_j}}  \sin \frac{\pi r}{N+1} \parens{\tilde \psi^\dagger_r + \tilde \psi^{}_r} \ket{k}.
\]
Combining this with
\[
    \sigma^+_n \sigma^-_n = \psi^\dagger_n \psi^{}_n = \frac 2{N+1} \sum_{ab} \sin \frac{\pi n a}{N+1} \sin \frac{\pi n b}{N+1}  \tilde \psi^\dagger_a \tilde \psi^{}_b,
\]
we can then calculate the dynamical expectation values
\begin{eqnarray*}
\fl \braopket{k}{\sigma^x_1 \expe^{\imi t H} \sigma^+_n \sigma^-_n \expe^{-\imi t H} \sigma^x_1}{k} 
        = \frac 2{N+1} \sum_{ab} \hat{\sum_{lr}} \expe^{- \imi t \parens{s_{k_r} \omega_r - s_{k_l} \omega_l}} \sin \frac{\pi n a}{N+1} \sin \frac{\pi n b}{N+1} \\
        \hspace{16em} \times \braopket{k}{\parens{\tilde \psi^\dagger_l + \tilde \psi^{}_l} \tilde \psi^\dagger_a \tilde \psi^{}_b \parens{\tilde \psi^\dagger_r + \tilde \psi^{}_r}}{k}.
\end{eqnarray*}

Again we use the anticommutation relations to calculate the matrix elements
\[
\fl \braopket{k}{\parens{\tilde \psi^\dagger_l + \tilde \psi^{}_l} \tilde \psi^\dagger_a \tilde \psi^{}_b \parens{\tilde \psi^\dagger_r + \tilde \psi^{}_r}}{k} 
        = \delta_{ab} \delta_{lr} k_a + \delta_{al} \delta_{br} (1 - k_a) (1 - k_b) - \delta_{ar} \delta_{bl} k_a k_b,
\]
yielding
\begin{eqnarray*} 
\fl \braopket{k}{\sigma^x_1 \expe^{\imi t H} \sigma^+_n \sigma^-_n \expe^{-\imi t H} \sigma^x_1}{k}
        = \frac 2{N+1} \sum_a \sin^2 \frac{\pi n a}{N+1} k_a + \frac 2{N+1} \hat{\sum_{ab}} \expe^{- \imi t (\omega_a - \omega_b)}\\
        \hspace{11em} \times \sin \frac{\pi n a}{N+1} \sin \frac{\pi n b}{N+1} \parens{(1 - k_a)(1 - k_b) - k_a k_b}.
\end{eqnarray*}
Since the first term
\[
    \frac 2{N+1} \sum_a \sin^2 \frac{\pi n a}{N+1} k_a = \braopket{k}{\sigma^+_n \sigma^-_n}{k}
\]
is just the local magnetization of the unperturbed state, we obtain eq.~(\ref{eq:eigenstate-spin-flip}).

\end{document}